\documentclass[letterpaper]{article} 
\usepackage{aaai24}  
\usepackage{times}  
\usepackage{helvet}  
\usepackage{courier}  
\usepackage[hyphens]{url}  
\usepackage{graphicx} 
\usepackage{natbib}  
\usepackage{caption} 
\usepackage{algorithm}
\usepackage{algorithmic}

\usepackage{graphicx}

\usepackage{multirow}
\usepackage{multicol}
\usepackage{booktabs}

\usepackage{newfloat}
\usepackage{listings}
\DeclareCaptionStyle{ruled}{labelfont=normalfont,labelsep=colon,strut=off} 
\floatstyle{ruled}
\newfloat{listing}{tb}{lst}{}
\floatname{listing}{Listing}
\title{Examining the Implications of Deepfakes for Election Integrity}
\author {
    Hriday Ranka\textsuperscript{1}\thanks{Equal contribution.},
    Mokshit Surana\textsuperscript{1}\footnotemark[1],
    Neel Kothari\textsuperscript{1}\footnotemark[1],
    Veer Pariawala\textsuperscript{1}\footnotemark[1],
    Pratyay Banerjee\textsuperscript{1}\footnotemark[1],
    Aditya Surve\textsuperscript{1}\footnotemark[1],
    Sainath Reddy Sankepally\textsuperscript{1}\footnotemark[1],
    Raghav Jain\textsuperscript{1},
    Jhagrut Lalwani\textsuperscript{1},
    Swapneel Mehta\textsuperscript{1}
}
\affiliations {
    \textsuperscript{1}SimPPL\
mokshitsurana3110@gmail.com
}

\usepackage{bibentry}

\begin{document}

\maketitle

\begin{abstract}
It is becoming cheaper to launch disinformation operations at scale using AI-generated content, in particular 'deepfake' technology. 
We have observed instances of deepfakes in political campaigns, where generated content is employed to both bolster the credibility of certain narratives (reinforcing outcomes) and manipulate public perception to the detriment of targeted candidates or causes (adversarial outcomes).
We discuss the threats from deepfakes in politics, highlight model specifications underlying different types of deepfake generation methods, and contribute an accessible evaluation of the efficacy of existing detection methods.
We provide this as a summary for lawmakers and civil society actors to understand how the technology may be applied in light of existing policies regulating its use.
We highlight the limitations of existing detection mechanisms and discuss the areas where policies and regulations are required to address the challenges of deepfakes.
\end{abstract}

\section{Introduction}

In recent years, the digital world has undergone rapid advancements, resulting in the emergence of sophisticated technologies that blur the boundaries between reality and fiction. 
During these events, deepfake technology has been a cause for concern due to its potential implications and dangerous consequences. 
Coined from the terms "deep learning" and "fake," deepfakes utilize advanced algorithms to generate hyper-realistic multimedia content, often indistinguishable from authentic material \cite{10.1145/3425780}.

While AI-generated images and deepfakes both make use of artificial intelligence, they serve vastly different purposes primarily arising from the intent behind content creation and distribution.
The former involves the creation of realistic images through algorithms that learn from real data, whereas the latter--technologically a subset thereof--typically aims to deceive viewers into believing they are accessing authentic content.
This article \cite{becker2023embracing} discusses the utilization of advanced technologies to produce lifelike and personalized dynamic facial visuals, as well as developing and adjusting various high-caliber static content.
However, \cite{wang2022deepfakes} underscores how intent matters, showcasing the potential abuse of deepfakes in producing fake images for scientific publications.

There are a variety of methods used to generate deepfakes; with state-of-the-art approaches including diffusion-based models and generative adversarial networks (GANs). 
Stable diffusion models, such as DALL-E 2, Midjourney, and Stable Diffusion are neural networks trained on a large dataset of images and captions to generate convincing images from text descriptions \cite{chen2023textimage}. 
On the other hand, GANs such as Cycle-GAN, DCGAN, and WGAN are deep learning systems commonly used for image generation, data augmentation, music generation, and deepfake creation \cite{remya2021detection}. 
The quality of deepfakes generated using GANs depends on the quantity and variety of training data, and the use of GANs to synthesize minimum training data for deepfake generation has been an area of active research \cite{singh2020using}.




\section{The Dangers of Deepfakes for Democratic Elections}

The use of deepfake technology to spread disinformation poses a significant threat to free and fair democratic elections. Deepfakes can serve as a potent tool for malicious actors to manipulate voters and influence election outcomes \cite{Bryandeepfakes2023,Appel2022deepfake}. There are several ways deepfakes endanger democratic processes:

\begin{enumerate}

\item Deepfakes can directly alter voter preferences and spread disinformation about candidates by making them appear to take policy positions they do not hold or engage in illegal behavior \cite{RAnd,Appel2022deepfake}. This could undermine trust in the electoral process.

\item Coordinated disinformation campaigns utilizing deepfake videos could prevent citizens from voting by spreading false information about election procedures or intimidating voters through blackmail \cite{pawelec2022deepfakes}. This form of voter suppression damages participation.

\item Deepfakes amplified through social media and messaging platforms can rapidly reach millions of viewers \cite{NChrist,CJ}, confusing them about candidates and issues. Widespread false or misleading information shaped by deepfakes harms informed civic discourse.

\item There are already instances of political deepfakes circulating globally, such as the manipulated videos of Nancy Pelosi and Volodymyr Zelenskyy \cite{CBS,JRhett}. In the 2023 Argentina elections, both leading candidates, Javier Milei and Sergio Massa, created and spread deepfake images and videos of each other to portray their opponent negatively \cite{DavidFeliba_2023}. This demonstrates the real-world vulnerability of elections.

\end{enumerate}

\section{DeepFake Creation and Identification}


\subsection{Generative Models}

Deepfakes are fueled by technological advancements in generative models, broadly including autoencoders, GANs, transformer-based models, and diffusion-based models.

%
The historical developments in deepfake technology are as follows \cite{masood2023deepfakes}:
\begin{itemize}
    \item \textbf{Pre-2014: Traditional Techniques and Early Autoencoders:} Before 2014, the field of manipulated multimedia predominantly utilized conventional methods such as splicing and copy-move, with an early occurrence dating as far back as 1860. Autoencoders, a generative model originating in the 1980s, garnered interest in the early 2000s and made significant contributions to the development of early generative models. Nevertheless, their influence on the progression of deepfake technology was diminished by more sophisticated models.
    \item \textbf{2014-2017: Emergence of GANs:} In 2014, the introduction of Generative Adversarial Networks (GANs) by Ian Goodfellow \cite{goodfellow2020generative} brought about a significant change in deepfake technology. During this time, GANs emerged as a highly influential and transformative factor, with academic initiatives such as Face2Face and Synthesizing Obama playing a significant role in the initial advancements. In September 2017, a significant event took place on Reddit when a user named "deepfake" shared the initial authentic deepfake. This entailed the utilization of computer-generated videos showcasing renowned actresses with their faces effectively replaced with explicit content. This occurrence garnered public interest, indicating a significant turning point in the advancement of deepfake technology. The event brought to light the fact that complex generative models could be used for bad and dishonest purposes, which raised awareness and led to more thought about ethical issues and regulatory actions.
    \item \textbf{2018 Onwards: Integration of Transformers and Diffusion Models:} Over the following years, the deepfake technology landscape continued to progress. The advent of open-source projects such as DeepFaceLab in 2018 has played a significant role in making deepfake creation tools more accessible. Furthermore, there was a significant change in the investigation of transformers beyond their original utilization in natural language processing. Researchers have acknowledged the adaptability of transformers, expanding their application to include image synthesis and other tasks unrelated to text.
\end{itemize}
Currently, the advancements in deepfake technology revolve around the integration of transformers and diffusion models. The objective of this collaborative approach is to attain generative outcomes of superior quality, with a focus on enhancing the authenticity and capabilities of the produced content. This technology advances with a focus on enhancing security. The progress is crucial, especially in the context of elections, where the threat of deepfakes contributing to misinformation campaigns continues to be a significant concern.


\subsection{GAN-based Architectures}

Very broadly, existing deepfake detection techniques can be divided into 2 categories, based on the consideration of change of characteristic/genuine attributes in space (spatial consideration) and the ones that consider changes in space as well as time (spatio-temporal). 
For the first class of techniques, researchers aim to capture spatial features and perform the classification of visual data (video/image) as real or fake \cite{bonettini2020video, app12199820, s21165413, Afchar_2018, Coccomini_2022}. 
On the other hand, models utilizing time-series and image features in combination have proven to be more accurate in identifying deepfakes \cite{delima2020deepfake, zhang2022deepfake, coccomini2022mintime, cai2023marlin, 9072088, tariq2020convolutional, cozzolino2021idreveal, wodajo2023deepfake}.

\subsubsection{Evaluating GAN-based DeepFake Generation}

\textbf{\\Accuracy Ratings}:
Prominent tools like FaceSwap-GAN \cite{An_2022} exhibit superior accuracy since perceptual loss improves the direction of eyeballs to be more realistic and consistent with the input face.
It also smoothes out artifacts in the segmentation mask, resulting in higher output quality and rendering dependable options for a range of uses.
Similarly, tools like Simswap, Fewshot FT Gan, and Faceshifter boast high accuracy.
But it's important to recognize that some tools—like FaceApp and StyleGAN—have poorer accuracy levels. 
StyleGAN \cite{Brownlee_2020} which relies on traditional GAN generators, inherits the interpretability and control issues associated with typical GAN models.
Due to the generators' limited knowledge of latent space qualities and image synthesis techniques, it may be difficult to comprehend and manipulate components inside the StyleGAN framework.

\textbf{Usability Analysis}:
CycleGAN\cite{zhu2017unpaired} stands out for its exceptional user-friendliness and versatility. 
Its simplicity lies in the elimination of the requirement for paired data (Paired training data consists of training examples {x i, y i } N i=1 , where the correspondence between x i and y i exists), making it accessible to a broader user base. 
The capability to seamlessly work with unpaired data not only simplifies the process but also proves cost-effective, addressing challenges associated with obtaining extensive and reliable paired datasets. 
This user-friendly approach, coupled with high accuracy in image-to-image translation, positions CycleGAN as a valuable tool.
The usability of face swapping and attribute manipulation tools like Faceapp, SimSwap, Fewshot FT GAN, and FaceShifter depends on factors such as user interface, documentation, and the level of technical expertise required.

\textbf{Security Assessment}:
By avoiding information loss, improving representation ability, and rejecting an attribute-independent constraint, AttGAN \cite{he2019attgan} prioritizes a secure facial attribute editing approach. 
The framework's security is strengthened by the attribute classification constraint on generated images, which guarantees accurate attribute manipulations. 
Utilizing adversarial learning and reconstruction provides additional resilience to maintain original facial features and produce realistic images. 
However, these implementation strategies have varied levels of defense against adversarial attacks, so they must be carefully considered. 
Overall, compared to models with more stringent constraints, AttGAN's combination of these features improves security

\textbf{Computational Efficiency}:
CycleGAN is recognized for its computational efficiency, especially in handling unpaired data.
Usability varies, with CycleGAN being user-friendly, while others, like Style-GAN variants, may demand deep learning expertise and substantial computational resources.





\subsection{Challenges faced by existing tools and techniques}
Challenges arise in the performance of deepfake detection algorithms when faced with low-quality films compared to high-resolution videos. 
Videos may undergo various transformations, including reshaping, rotations, and compression, necessitating flexible detection algorithms to maintain efficacy.

Time consumption emerged as a significant concern for real-world applications of deepfake-detection techniques. Despite their potential impact on social security, existing detection methods still face limitations in terms of extensive time requirements, hindering widespread adoption in practical scenarios.

The challenge of insufficient data for specific characters during the creation of deepfake models was highlighted. While models are often trained on specific datasets, they may struggle to produce accurate outputs when faced with limited data for a particular character. Retraining models for each distinct target character is a time-consuming process.

Dataset quality was identified as another challenging area, with most datasets created under ideal conditions that differ from real-world testing scenarios. This misalignment in dataset quality adds complexity to the development and evaluation of deepfake-detection algorithms.

Despite the availability of various deepfake-generation tools, inherent flaws and limitations persist. 
These tools are often tailored to specific traits, emphasizing the need for additional research to enhance their efficiency. 
Consequently, the creation of general-purpose deepfake-generation tools remains a complex and challenging process that warrants further investigation.

\section{Reviewing Platform Policies against Deepfakes}

The regulation of deepfake content on social media platforms has become a critical issue in recent years. 

With the rise of AI-generated manipulated media, platforms like Meta, X (formerly Twitter), Reddit, Tiktok, and YouTube have implemented various policies to address the spread of deepfakes \cite{policies}.

\subsection{Meta} \cite{metappolicy}

\begin{itemize}
\item Removal of Manipulated Media: Meta will remove audio, photos, or videos, including deepfakes, if they violate any of their Community Standards, such as those related to graphic violence, voter suppression, and hate speech.

\item Detection Efforts: Meta has launched the Deep Fake Detection Challenge and is collaborating with experts to address deepfakes and manipulated media.  
\end{itemize}

\subsection{X (formerly Twitter)} \cite{xppolicy}

\begin{itemize}
\item Prohibition of Misleading Media: X prohibits the sharing of synthetic, manipulated, or out-of-context media that may deceive or confuse people and lead to harm.  

However, memes, satire, animations, and cartoons are generally not in violation of this policy.

\item Labeling and Consequences: In some cases, X may label posts containing misleading media and take action to reduce the visibility of the post on the platform.
\end{itemize}

\subsection{YouTube} \cite{youtubeppolicy}

\begin{itemize}  
\item Disclosure Requirement: YouTube will require creators to disclose altered or synthetic content that is realistic, including using AI tools. The platform will inform viewers about such content through labels in the description panel and video player.
\item Removal and Labeling: YouTube may remove AI-generated or manipulated content that simulates an identifiable individual, and it will work with creators to ensure they understand the new requirements.
\end{itemize}

\subsection{Reddit} \cite{redditpolicy1}

\begin{itemize}
    \item Reddit does not allow content that impersonates individuals or entities in a misleading or deceptive manner, including deepfakes or other manipulated content presented to mislead, or falsely attributed to an individual or entity. 
\end{itemize}

\subsection{Tiktok} \cite{tiktokpolicy1,tiktokpolicy2}

\begin{itemize}
    \item TikTok bans deepfakes of private figures and young people, and all realistic AI deepfakes must be "clearly disclosed".
\end{itemize}

\subsection{Monitoring the Implementation of Current Deepfake Regulations}

Monitoring the enforcement of current deepfake regulations involves assessing the adherence of various stakeholders, evaluating the efficiency of detection tools, and addressing emerging challenges. 
We explore the strategies and considerations involved in overseeing the implementation of existing deepfake regulations:

\begin{enumerate}
    \item \textbf{Stakeholder Compliance:} Regulatory success hinges on the compliance of key stakeholders, including social media platforms, technology companies, political campaigns, and content creators. 
    Continuous monitoring of these entities is essential to ensuring they are actively adopting measures to prevent the creation and dissemination of malicious deepfakes. 
    Collaborative efforts between regulatory bodies and stakeholders can facilitate the exchange of best practices, ensuring a unified approach to tackling deepfake-related threats.
    
    \item \textbf{Effectiveness of Detection Tools:} The efficacy of deepfake detection tools plays a pivotal role in enforcing regulations. 
    Regular assessments of the performance of existing detection mechanisms are crucial to identifying strengths, weaknesses, and areas for improvement. 
    This involves evaluating the accuracy, speed, and adaptability of tools for detecting evolving deepfake techniques. 
    Ongoing research and development are necessary to enhance the capabilities of detection tools and address emerging challenges in real time.

    \item \textbf{Educational Initiatives:} Monitoring the implementation of regulations extends beyond enforcement measures to include educational initiatives. 
    Informing the public, political candidates, and election officials about the existence of deepfake threats, the regulatory framework in place, and preventive measures is vital. Periodic assessments of the effectiveness of educational campaigns can guide adjustments and refinements to ensure they remain relevant and impactful.

    \item \textbf{Adaptability to Evolving Threats:} The landscape of deepfake technology is dynamic, with new advancements and variations emerging regularly. 
    Monitoring the implementation of regulations requires a proactive approach to stay ahead of evolving threats. 
    Regulatory bodies should establish mechanisms for continuous threat assessment, allowing timely modification of regulations to address emerging challenges effectively.

    \item \textbf{International Collaboration:} Given the global nature of information dissemination and potential cross-border impact, international collaboration is essential for effective regulation. 
    Monitoring the implementation of deepfake regulations involves fostering partnerships between countries, sharing intelligence, and collectively addressing challenges. 
    Regular forums for collaboration can facilitate the exchange of insights and strategies to combat the transnational aspects of deepfake threats.
\end{enumerate}

\subsection{Areas where interventions/policies and regulations are required}
The emergence of deepfakes poses significant challenges to the integrity of elections. To address these challenges, various areas require targeted interventions, policies, and regulations.

\begin{enumerate}
    \item \textbf{Authentication Protocols:} Implementing protocols to authenticate digital content can help distinguish genuine media from deepfakes. 
    This includes the development of digital watermarks or certification systems \cite{westerlund2019emergence}.

    \item \textbf{Transparency Requirements:} Legislation mandating the disclosure of AI-manipulated content can increase transparency. 
    Any altered media should be clearly labeled to inform the public about its modified nature \cite{langa2021deepfakes}.

    \item \textbf{Media Literacy Programs:} Educating the public, especially voters, about the existence and nature of deepfakes is crucial. 
    Media literacy programs can teach people how to critically assess and verify the authenticity of the information they receive \cite{el2023effect}.

    \item \textbf{Legal Frameworks Against Misuse:} There should be clear legal consequences for maliciously creating or distributing deepfakes \cite{feeney2021deepfake}.
\end{enumerate}

\section{Conclusion}
This paper urgently calls for societal, policymaker, and regulatory action to safeguard elections from the pervasive threat of deepfake technology. It meticulously explores the dissemination of deepfakes and their implications for election security, advocating for comprehensive regulatory frameworks to counter the growing accessibility of AI-generated disinformation. The paper delves into the intricacies of the deepfake generation, emphasizing its profound impact on politics, trust, and democratic processes. While detection tools show promise, challenges persist, necessitating research and adaptability. Varied deepfake policies among technology and social media companies underscore the importance of vigilant monitoring for effective implementation. Recognizing the dynamic nature of these challenges, the paper urges continuous proactive efforts from policymakers, technologists, and the public to enhance detection capabilities and safeguard democracy against the evolving threats of deepfake technology. Ongoing commitment to innovation, collaboration, and democratic principles is crucial for ensuring the resilience of electoral processes against AI-generated disinformation.
We also express our sincere gratitude to Mr. Eric Davis for his invaluable feedback and insightful comments on this paper.

\bibliography{CameraReady/LaTeX/aaai24}

\begin{thebibliography}{49}
\providecommand{\natexlab}[1]{#1}

\bibitem[{Afchar et~al.(2018)Afchar, Nozick, Yamagishi, and Echizen}]{Afchar_2018}
Afchar, D.; Nozick, V.; Yamagishi, J.; and Echizen, I. 2018.
\newblock MesoNet: a Compact Facial Video Forgery Detection Network.
\newblock In \emph{2018 IEEE International Workshop on Information Forensics and Security (WIFS)}. IEEE.

\bibitem[{An(2022)}]{An_2022}
An, S. 2022.
\newblock Shaoanlu/faceswap-gan: A denoising autoencoder + adversarial losses and attention mechanisms for face swapping.

\bibitem[{Appel and Prietzel(2022)}]{Appel2022deepfake}
Appel, M.; and Prietzel, F. 2022.
\newblock {The detection of political deepfakes}.
\newblock \emph{Journal of Computer-Mediated Communication}, 27(4): zmac008.

\bibitem[{Becker and Laycock(2023)}]{becker2023embracing}
Becker, C.; and Laycock, R. 2023.
\newblock Embracing deepfakes and AI-generated images in neuroscience research.
\newblock \emph{European Journal of Neuroscience}.

\bibitem[{Bonettini et~al.(2020)Bonettini, Cannas, Mandelli, Bondi, Bestagini, and Tubaro}]{bonettini2020video}
Bonettini, N.; Cannas, E.~D.; Mandelli, S.; Bondi, L.; Bestagini, P.; and Tubaro, S. 2020.
\newblock Video Face Manipulation Detection Through Ensemble of CNNs.
\newblock arXiv:2004.07676.

\bibitem[{Brownlee(2020)}]{Brownlee_2020}
Brownlee, J. 2020.
\newblock A gentle introduction to stylegan the style generative Adversarial Network.

\bibitem[{Cai et~al.(2023)Cai, Ghosh, Stefanov, Dhall, Cai, Rezatofighi, Haffari, and Hayat}]{cai2023marlin}
Cai, Z.; Ghosh, S.; Stefanov, K.; Dhall, A.; Cai, J.; Rezatofighi, H.; Haffari, R.; and Hayat, M. 2023.
\newblock MARLIN: Masked Autoencoder for facial video Representation LearnINg.
\newblock arXiv:2211.06627.

\bibitem[{Center(2023)}]{policies}
Center, S. U.~D. 2023.
\newblock {Social Media Policies: Mis/Disinformation, Threats, and Harassment}.
\newblock \url{https://statesuniteddemocracy.org/resources/social-media-policies/}.
\newblock [Online; accessed 30-November-2023].

\bibitem[{Chen et~al.(2023)Chen, Haldar, Akhtar, and Mian}]{chen2023textimage}
Chen, Y.; Haldar, N. A.~H.; Akhtar, N.; and Mian, A. 2023.
\newblock Text-image guided Diffusion Model for generating Deepfake celebrity interactions.
\newblock arXiv:2309.14751.

\bibitem[{Christopher(2020)}]{NChrist}
Christopher, N. 2020.
\newblock {We've Just Seen the First Use of Deepfakes in an Indian Election Campaign}.
\newblock \url{https://www.vice.com/en/article/jgedjb/the-first-use-of-deepfakes-in-indian-election-by-bjp}.
\newblock [Online; accessed 21-November-2023].

\bibitem[{Coccomini et~al.(2022{\natexlab{a}})Coccomini, Messina, Gennaro, and Falchi}]{Coccomini_2022}
Coccomini, D.~A.; Messina, N.; Gennaro, C.; and Falchi, F. 2022{\natexlab{a}}.
\newblock \emph{Combining EfficientNet and Vision Transformers for Video Deepfake Detection}, 219–229.
\newblock Springer International Publishing.
\newblock ISBN 9783031064333.

\bibitem[{Coccomini et~al.(2022{\natexlab{b}})Coccomini, Zilos, Amato, Caldelli, Falchi, Papadopoulos, and Gennaro}]{coccomini2022mintime}
Coccomini, D.~A.; Zilos, G.~K.; Amato, G.; Caldelli, R.; Falchi, F.; Papadopoulos, S.; and Gennaro, C. 2022{\natexlab{b}}.
\newblock MINTIME: Multi-Identity Size-Invariant Video Deepfake Detection.
\newblock arXiv:2211.10996.

\bibitem[{Cozzolino et~al.(2021)Cozzolino, Rössler, Thies, Nießner, and Verdoliva}]{cozzolino2021idreveal}
Cozzolino, D.; Rössler, A.; Thies, J.; Nießner, M.; and Verdoliva, L. 2021.
\newblock ID-Reveal: Identity-aware DeepFake Video Detection.
\newblock arXiv:2012.02512.

\bibitem[{David~Feliba(2023)}]{DavidFeliba_2023}
David~Feliba, T. R.~F. 2023.
\newblock How Argentina’s new president-elect used AI to target his opponents during the campaign.

\bibitem[{de~Lima et~al.(2020)de~Lima, Franklin, Basu, Karwoski, and George}]{delima2020deepfake}
de~Lima, O.; Franklin, S.; Basu, S.; Karwoski, B.; and George, A. 2020.
\newblock Deepfake Detection using Spatiotemporal Convolutional Networks.
\newblock arXiv:2006.14749.

\bibitem[{DeepTraceTechnologies(2023)}]{DeepTrace_Technologies}
DeepTraceTechnologies. 2023.
\newblock "\url{https://www.deeptracetech.com/}.
\newblock [Online; accessed 25-November-2023].

\bibitem[{El~Mokadem(2023)}]{el2023effect}
El~Mokadem, S.~S. 2023.
\newblock The Effect of Media Literacy on Misinformation and Deep Fake Video Detection.
\newblock \emph{Arab Media \& Society}, 35.

\bibitem[{Feeney(2021)}]{feeney2021deepfake}
Feeney, M. 2021.
\newblock Deepfake Laws Risk Creating More Problems Than They Solve.
\newblock \emph{Regulatory Transparency Project}.

\bibitem[{Goldstein and DiResta(2022)}]{goldstein2022research}
Goldstein, J.~A.; and DiResta, R. 2022.
\newblock Research Note: This Salesperson Does Not Exist: How Tactics from Political Influence Operations on Social Media are Deployed for Commercial Lead Generation.
\newblock \emph{Harvard Kennedy School Misinformation Review}, 3(5): 1--15.

\bibitem[{Goodfellow et~al.(2020)Goodfellow, Pouget-Abadie, Mirza, Xu, Warde-Farley, Ozair, Courville, and Bengio}]{goodfellow2020generative}
Goodfellow, I.; Pouget-Abadie, J.; Mirza, M.; Xu, B.; Warde-Farley, D.; Ozair, S.; Courville, A.; and Bengio, Y. 2020.
\newblock Generative adversarial networks.
\newblock \emph{Communications of the ACM}, 63(11): 139--144.

\bibitem[{Hasan and Salah(2019)}]{hasan2019combating}
Hasan, H.~R.; and Salah, K. 2019.
\newblock Combating deepfake videos using blockchain and smart contracts.
\newblock \emph{Ieee Access}, 7: 41596--41606.

\bibitem[{He et~al.(2019)He, Zuo, Kan, Shan, and Chen}]{he2019attgan}
He, Z.; Zuo, W.; Kan, M.; Shan, S.; and Chen, X. 2019.
\newblock Attgan: Facial attribute editing by only changing what you want.
\newblock \emph{IEEE transactions on image processing}, 28(11): 5464--5478.

\bibitem[{Ismail et~al.(2021)Ismail, Elpeltagy, S.~Zaki, and Eldahshan}]{s21165413}
Ismail, A.; Elpeltagy, M.; S.~Zaki, M.; and Eldahshan, K. 2021.
\newblock A New Deep Learning-Based Methodology for Video Deepfake Detection Using XGBoost.
\newblock \emph{Sensors}, 21(16).

\bibitem[{Jee(2020)}]{CJ}
Jee, C. 2020.
\newblock {An Indian politician is using deepfake technology to win new voters}.
\newblock \url{https://www.technologyreview.com/2020/02/19/868173/an-indian-politician-is-using-deepfakes-to-try-and-win-voters/}.
\newblock [Online; accessed 30-November-2023].

\bibitem[{Jung, Kim, and Kim(2020)}]{9072088}
Jung, T.; Kim, S.; and Kim, K. 2020.
\newblock DeepVision: Deepfakes Detection Using Human Eye Blinking Pattern.
\newblock \emph{IEEE Access}, 8: 83144--83154.

\bibitem[{Langa(2021)}]{langa2021deepfakes}
Langa, J. 2021.
\newblock Deepfakes, real consequences: Crafting legislation to combat threats posed by deepfakes.
\newblock \emph{BUL Rev.}, 101: 761.

\bibitem[{Masood et~al.(2023)Masood, Nawaz, Malik, Javed, Irtaza, and Malik}]{masood2023deepfakes}
Masood, M.; Nawaz, M.; Malik, K.~M.; Javed, A.; Irtaza, A.; and Malik, H. 2023.
\newblock Deepfakes generation and detection: State-of-the-art, open challenges, countermeasures, and way forward.
\newblock \emph{Applied intelligence}, 53(4): 3974--4026.

\bibitem[{McKenzie(2023)}]{Bryandeepfakes2023}
McKenzie, B. 2023.
\newblock {IS THAT REAL? DEEPFAKES COULD POSE DANGER TO FREE ELECTIONS}.
\newblock \url{https://news.virginia.edu/content/real-deepfakes-could-pose-danger-free-elections}.
\newblock [Online; accessed 20-November-2023].

\bibitem[{Miller(2022)}]{JRhett}
Miller, J.~R. 2022.
\newblock {Deepfake video of Zelensky telling Ukrainians to surrender removed from social platforms}.
\newblock \url{https://nypost.com/2022/03/17/deepfake-video-shows-volodymyr-zelensky-telling-ukrainians-to-surrender/}.
\newblock [Online; accessed 21-November-2023].

\bibitem[{Mirsky and Lee(2021)}]{10.1145/3425780}
Mirsky, Y.; and Lee, W. 2021.
\newblock The Creation and Detection of Deepfakes: A Survey.
\newblock \emph{ACM Comput. Surv.}, 54(1).

\bibitem[{Monika~Bickert(2020)}]{metappolicy}
Monika~Bickert, G. P.~M., Vice~President. 2020.
\newblock {Enforcing Against Manipulated Media}.
\newblock \url{https://about.fb.com/news/2020/01/enforcing-against-manipulated-media/}.
\newblock [Online; accessed 30-November-2023].

\bibitem[{NBC-News(2023)}]{tiktokpolicy1}
NBC-News. 2023.
\newblock {TikTok bans deepfakes of young people as it updates guidelines}.
\newblock \url{https://www.nbcnews.com/tech/tech-news/tiktok-bans-deepfakes-young-people-updates-guidelines-rcna75949}.
\newblock [Online; accessed 30-November-2023].

\bibitem[{News(2019)}]{CBS}
News, C. 2019.
\newblock {Doctored Nancy Pelosi video highlights threat of "deepfake" tech}.
\newblock \url{https://www.cbsnews.com/news/doctored-nancy-pelosi-video-highlights-threat-of-deepfake-tech-2019-05-25/}.
\newblock [Online; accessed 21-November-2023].

\bibitem[{Pawelec(2022)}]{pawelec2022deepfakes}
Pawelec, M. 2022.
\newblock Deepfakes and democracy (theory): how synthetic audio-visual media for disinformation and hate speech threaten core democratic functions.
\newblock \emph{Digital society}, 1(2): 19.

\bibitem[{Peters(2020)}]{redditpolicy1}
Peters, J. 2020.
\newblock {Reddit bans impersonation on its platform}.
\newblock \url{https://www.theverge.com/2020/1/9/21058803/reddit-account-ban-impersonation-policy-deepfakes-satire-rules}.
\newblock [Online; accessed 30-November-2023].

\bibitem[{PTI-News(2023)}]{youtubeppolicy}
PTI-News. 2023.
\newblock YouTube Users Have To Disclose Altered Content That Looks Realistic: Google.
\newblock \url{https://www.bqprime.com/nation/youtube-users-have-to-disclose-altered-content-that-looks-realistic-google}.
\newblock [Online; accessed 30-November-2023].

\bibitem[{Ray(2021)}]{RAnd}
Ray, A. 2021.
\newblock Disinformation, deepfakes and democracies: The need for legislative reform.
\newblock \emph{The University of New South Wales Law Journal}, 44(3): 983–1013.

\bibitem[{Raza, Munir, and Almutairi(2022)}]{app12199820}
Raza, A.; Munir, K.; and Almutairi, M. 2022.
\newblock A Novel Deep Learning Approach for Deepfake Image Detection.
\newblock \emph{Applied Sciences}, 12(19).

\bibitem[{Remya~Revi, Vidya, and Wilscy(2021)}]{remya2021detection}
Remya~Revi, K.; Vidya, K.; and Wilscy, M. 2021.
\newblock Detection of Deepfake Images Created Using Generative Adversarial Networks: A Review.
\newblock In \emph{Second International Conference on Networks and Advances in Computational Technologies: NetACT 19}, 25--35. Springer.

\bibitem[{SensityAI(2023)}]{Sensity_2023}
SensityAI. 2023.
\newblock \url{https://sensity.ai/}.
\newblock [Online; accessed 25-November-2023].

\bibitem[{Singh, Sharma, and Smeaton(2020)}]{singh2020using}
Singh, S.; Sharma, R.; and Smeaton, A.~F. 2020.
\newblock Using GANs to Synthesise Minimum Training Data for Deepfake Generation.
\newblock arXiv:2011.05421.

\bibitem[{Tariq, Lee, and Woo(2020)}]{tariq2020convolutional}
Tariq, S.; Lee, S.; and Woo, S.~S. 2020.
\newblock A Convolutional LSTM based Residual Network for Deepfake Video Detection.
\newblock arXiv:2009.07480.

\bibitem[{Vincent(2023)}]{tiktokpolicy2}
Vincent, J. 2023.
\newblock {TikTok bans deepfakes of nonpublic figures and fake endorsements in rule refresh}.
\newblock \url{https://www.theverge.com/2023/3/21/23648099/tiktok-content-moderation-rules-deepfakes-ai}.
\newblock [Online; accessed 30-November-2023].

\bibitem[{Wang et~al.(2022)Wang, Zhou, Yang, and Yu}]{wang2022deepfakes}
Wang, L.; Zhou, L.; Yang, W.; and Yu, R. 2022.
\newblock Deepfakes: a new threat to image fabrication in scientific publications?
\newblock \emph{Patterns}, 3(5).

\bibitem[{Westerlund(2019)}]{westerlund2019emergence}
Westerlund, M. 2019.
\newblock The emergence of deepfake technology: A review.
\newblock \emph{Technology innovation management review}, 9(11).

\bibitem[{Wodajo, Atnafu, and Akhtar(2023)}]{wodajo2023deepfake}
Wodajo, D.; Atnafu, S.; and Akhtar, Z. 2023.
\newblock Deepfake Video Detection Using Generative Convolutional Vision Transformer.
\newblock arXiv:2307.07036.

\bibitem[{X(2023)}]{xppolicy}
X. 2023.
\newblock {Synthetic and manipulated media policy}.
\newblock \url{https://help.twitter.com/en/rules-and-policies/manipulated-media}.
\newblock [Online; accessed 30-November-2023].

\bibitem[{Zhang et~al.(2022)Zhang, Lin, Hua, Wang, Zeng, and Ge}]{zhang2022deepfake}
Zhang, D.; Lin, F.; Hua, Y.; Wang, P.; Zeng, D.; and Ge, S. 2022.
\newblock Deepfake Video Detection with Spatiotemporal Dropout Transformer.
\newblock arXiv:2207.06612.

\bibitem[{Zhu et~al.(2017)Zhu, Park, Isola, and Efros}]{zhu2017unpaired}
Zhu, J.-Y.; Park, T.; Isola, P.; and Efros, A.~A. 2017.
\newblock Unpaired image-to-image translation using cycle-consistent adversarial networks.
\newblock In \emph{Proceedings of the IEEE international conference on computer vision}, 2223--2232.

\end{thebibliography}

\section{Appendix}\label{apd:first}

\section{From Theory to Practice: Deepfake Detection Examples}
Analyzing the effectiveness of deepfake-detection tools. 

1. \textbf{Sensity AI}\cite{Sensity_2023}:
Sensity AI presents a method for identifying GAN-generated images, especially those that belong to the StyleGAN2 model. 
Sensity's model was used to analyze 975 profile pictures from LinkedIn accounts. 
The research study focused on fictitious accounts on the platform with GAN-generated profile photos \cite{goldstein2022research}. 
As evidenced by the results, there was over 90\% confidence in the ability to identify GAN-generated images for 968 profile pictures. 
Surprisingly, 900 of these images had a confidence score higher than 99.9\%, demonstrating the resilience of Sensity's model. 
The majority of the identified images were linked to the StyleGAN2 model, highlighting the efficacy of Sensity AI in detecting specific types of GAN-generated content.

2. \textbf{Truepic}:
A startup company based in the United States has created a system that utilizes mobile apps to allow regular users and freelancers to capture images and store them on the company's servers.
The purpose of saving the images is to maintain their integrity. As a result, comparing any fraudulent attempt with the image kept on the servers makes it easy to spot.
Truepic utilizes blockchain technology \cite{hasan2019combating} to securely store metadata associated with saved images, guaranteeing their immutability. 
This method is highly dependent on placing a significant amount of trust in Truepic about the authenticity and integrity of the uploaded images. 
The operational details of incorporating logos, text tickers, subtitles, or closed captions into images or video frames are not readily apparent.

3. \textbf{Deeptrace}\cite{DeepTrace_Technologies}:
Deeptrace employs machine learning algorithms for detecting deepfake videos, showcasing high accuracy rates achieved through comprehensive audio-based, visual-based, and text-based analysis. 
Its real-time detection capabilities, scalability for analyzing large datasets, and continuous learning to adapt to new deepfake techniques position it as a versatile solution. 
Deeptrace effectively addresses the challenges posed by deepfakes, making it a valuable asset for organizations seeking reliable detection tools.

\end{document}